\documentclass[11pt]{article}
\usepackage{a4wide}
\usepackage{latexsym}
\usepackage[psamsfonts]{amssymb}
\usepackage{euscript}
\usepackage[psamsfonts]{amsfonts}

\newcommand{\bpXst}{\widetilde{X_{\star}}}
\newcommand{\Comi}[2]{\mathbf{C}_{i}(#1 , ( #2) )}
\newcommand{\Commi}[1]{\mathbf{C}_{i}(#1 )}
\newcommand{\con}{\ensuremath{\mathtt{con}}}
\newcommand{\weak}{\ensuremath{\mathtt{weak}}}
\newcommand{\der}{\ensuremath{\mathtt{der}}}
\newcommand{\promote}[1]{#1^{\dagger}}
\newcommand{\Ap}{\ensuremath{\mathtt{Ap}}}
\newcommand{\linimpl}{- \!\! \circ \,}
\newcommand{\llbang}{\, !}
\newcommand{\termobject}{\ensuremath{\mathsf{1}}}
\newcommand{\tarrow}{\ensuremath{\mathsf{t}}}
\newcommand{\isoarrow}{\stackrel{ \cong}{\longrightarrow}}
\newcommand{\CChom}[2]{\EuScript{C}( #1 , #2 )}
\newcommand{\CCshom}[2]{\EuScript{C}_{s}( #1 , #2 )}
\newcommand{\CC}{\EuScript{C}}
\newcommand{\CCs}{\EuScript{C}_{s}}
\newcommand{\CCt}[2]{\EuScript{C}_{t}( #1 , #2 )}
\newcommand{\bp}[1]{#1_{\star}}
\newcommand{\Cpo}{\ensuremath{\mathbf{Cpo}}}
\newcommand{\ident}{\ensuremath{\mathtt{id}}}
\newcommand{\Setst}{\ensuremath{\bp{\mathbf{Set}}}}
\newcommand{\Set}{\ensuremath{\mathbf{Set}}}
\newcommand{\Nat}{\ensuremath{\mathbb{N}}}
\newcommand{\nat}{\ensuremath{\mathbf{nat}}}
\newcommand{\casek}{\ensuremath{\mathtt{case}_{k}}}
\newcommand{\casel}{\ensuremath{\mathtt{case}_{l}}}
\newcommand{\YY}{\ensuremath{\mathbf{Y}}}
\newcommand{\Converges}{{\Downarrow}}
\newcommand{\Trm}[2]{\mathsf{Trm}( #1 , #2 )}
\newcommand{\Ctxt}[2]{\mathsf{Ctxt}( #1 , #2 )}
\newlength{\sqpreordheight}
\newlength{\sqpreorddepth}
\newcommand{\sqpreord}%
{\mathbin{\raisebox{-1.02ex}[\sqpreordheight][\sqpreorddepth]%
{$\stackrel{\textstyle \sqsubset}{\sim}$}}}
\newcommand{\sqgtpreord}%
{\mathbin{\raisebox{-1.02ex}[\sqpreordheight][\sqpreorddepth]%
{$\stackrel{\textstyle \sqsupset}{\sim}$}}}

\newcommand{\ipreord}{\lesssim}
\newcommand{\preord}{\sqpreord}
\newcommand{\Natst}{\widetilde{\bp{\Nat}}}
\newcommand{\fix}[1]{ #1^{\nabla}}
\newcommand{\lsem}{[\! [}

\newcommand{\Coh}{\mathbf{Coh}}
\newcommand{\rsem}{]\! ]}
\newcommand{\KK}{\EuScript{K}}
\newcommand{\KCC}{K_{!}(\CC )}
\newcommand{\norm}[1]{ \| #1 \|}
\newcommand{\CCpd}{\EuScript{C}^{\pi}}

\newcommand{\CCwo}{\EuScript{C}^{wo}}
\newcommand{\CCtwo}{\EuScript{C}_{t}^{wo}}
\newcommand{\CCtwohom}[2]{\EuScript{C}_{t}^{wo}( #1 , #2 )}
\newcommand{\proof}{\textsc{Proof}}
\newcommand{\qed}{\Box}
\newtheorem{theorem}{Theorem}[section]
\newtheorem{lemma}{Lemma}[section]
\newtheorem{proposition}{Proposition}[section]
\newcommand{\name}[1]{\ulcorner #1 \urcorner}

\begin{document}

\bibliographystyle{alpha}

\title{Axioms for Definability and Full Completeness}
\author{Samson Abramsky
}

\maketitle

\begin{abstract}
	Axioms are presented on models of PCF from which full abstraction 
	can be proved. These axioms have been distilled from recent results 
	on definability and full abstraction of game semantics for a number 
	of programming languages. Full completeness for pure simply-typed 
	$\lambda$-calculus is also axiomatized.
\end{abstract}

\section{Introduction}
The term ``full abstraction'' was coined by Robin Milner in 
\cite{Mil75}. In 1977 two seminal papers on the programming language 
PCF, by Milner and Plotkin, appeared in \textsl{Theoretical Computer Science} 
\cite{Mil77,Plo77}. These papers initiated an extensive body of work 
centering on the Full Abstraction problem for PCF 
(see \cite{BCL85,Cur93,Ong95} for surveys).

The importance of full abstraction for the semantics of programming 
languages is that it is one of the few quality filters we have. 
Specifically, it provides a clear criterion for assessing how 
definitive a semantic analysis of some language is. It must be 
admitted that to date the quest for fully abstract models has not 
yielded many obvious applications; but it has generated much of the 
deepest work in semantics. Perhaps it is early days yet.

Recently, game semantics has been used to give the first 
syntax-independent constructions of fully abstract models for a number 
of programming languages, including PCF  \cite{AJM,HO,Nic94},
richer functional languages \cite{AM95,M96a,M96b,HY97}, and
languages with 
non-functional features such as reference types and non-local control 
constructs \cite{AM97a,AM97b,AM97c,Lai97}. A noteworthy feature is that the key 
definability results 
for the richer languages are proved by a reduction to definability for 
the functional fragment, using a technique of \emph{factorization 
theorems}. Thus the results originally obtained for PCF prove to be a 
lynch-pin in the analysis of a much wider class of languages. 

When some success has 
been achieved with concrete constructions, it becomes important to 
identify the key properties of these constructions at a more abstract 
level; thus the trend towards axiomatic and synthetic domain theory, 
for example \cite{Fio96,Hyl91}. There has also been considerable progress in 
axiomatizing sufficient conditions for computational adequacy 
\cite{FP94,Brau96,M96b}. 
In another vein, the work on action structures \cite{MMP95} can be seen as 
an axiomatics for process calculi and other computational formalisms.

In the present paper we make the first contribution towards an 
axiomatic account of full abstraction, with specific reference to PCF. 
We present axioms on models of PCF from which the key results on 
definability and full abstraction can be proved. It should be 
emphasized that not only the results of \cite{AJM}, but also the 
top-level structure of the actual proofs, are captured by our 
axiomatic account. In particular, our main axioms are abstracted from 
key lemmas in \cite{AJM}. The axioms mostly take the form of 
assertions that some canonical map is an isomorphism, which is quite 
standard in categorical axiomatizations, for example of Synthetic 
Differential Geometry \cite{Koc81}, distributive categories 
\cite{Wal92}, or dualizing objects \cite{MMO91}.
It is also 
noteworthy that, although our results apply to intuitionistic types, the axioms 
make essential use of the linear decompositions of these types 
\cite{Gir87}.

The present paper is only a first step. We hope that it will lead to 
further work in a number of directions:
\begin{itemize}
	\item  a more abstract perspective on game semantics
	\item  a synthetic/axiomatic account of sequentiality
	\item  general results on full abstraction
	\item a fine structure theory of ``behavioural features'' of 
	categorical models.
	
	\end{itemize}
	
\section{Preliminaries}

This section is concerned with setting up some basic language in which 
the substantive axioms can be expressed.

\subsection{Affine categories}
We firstly recall the standard definition of a categorical model for 
the ($\otimes$, $\multimap$, $\times$, $\termobject$, $\llbang$) fragment of 
Affine logic \cite{See87,Bie95}.

An \emph{affine category} is a symmetric monoidal closed category 
$\CC$ 
with finite products, such that the tensor unit is the terminal 
object $\termobject$, together with a comonad which we write in 
``co-Kleisli form'' \cite{Man76} as $(\llbang , \der, \promote{(\cdot )})$ where
$ \llbang  : \CC \rightarrow \CC$, $\der_{A} : \llbang A \rightarrow A$ for all $A$, 
and 
\[ \promote{(\cdot )}_{A,B} : \CC ( \llbang A , B) \rightarrow \CC ( \llbang A , 
\llbang B ) \]
satisfy:
\begin{eqnarray}
	f^{\dagger} ; g^{\dagger} & = & (f^{\dagger} ; g)^{\dagger} \\
	f^{\dagger} ; \der & = & f \\
	\der^{\dagger} ; f & = & f .
\end{eqnarray}
There are moreover natural isomorphisms (the ``exponential laws'' )
\[ \begin{array}{ccc}
e_{A,B} & : & \llbang (A \times B) \isoarrow \llbang A \otimes \llbang B \\
e_{\termobject} & : & \llbang \termobject \isoarrow \termobject 
\end{array} \]
satisfying the coherence conditions given in \cite{Bie95}. Every object 
$\llbang A$ has a cocommutative comonoid structure given by
\begin{displaymath}
	\con_{A} = 
	\llbang A \stackrel{\llbang \Delta}{\longrightarrow}  \llbang  (A \times A) 
	\stackrel{e_{A,A}}{\longrightarrow}  \llbang A 
	\otimes \llbang A
\end{displaymath}
\begin{displaymath}
\weak_{A} = 
\llbang A \stackrel{\llbang \tarrow_{A}}{\longrightarrow}  \llbang \termobject  
\stackrel{e_{\termobject}}{\longrightarrow}  \termobject .
\end{displaymath}
The co-Kleisli category $K_{\llbang }(\CC )$ is cartesian closed, with 
products as in $\CC$, and function space 
\[ A \Rightarrow B  = \llbang A \linimpl B. \]

\subsection{Partiality}
Recall firstly that $\Setst$, the category of \emph{pointed sets}, has 
as objects sets with a specified base point, and as morphisms 
functions preserving the base point. $\Setst$ is symmetric monoidal 
closed, with tensor given by smash product and function spaces by 
basepoint-preserving maps. It has products as in $\Set$, and 
coproducts given by coalesced sums (\textit{i.e.} disjoint unions 
with basepoints identified). We write $\coprod$ for disjoint union of 
sets, and $+$ for coproduct in $\Setst$. We write $\bp{X}$ for an 
object of $\Setst$, where $X$ is the set of elements excluding the 
basepoint. Thus $\bp{X} + \bp{Y} = \bp{(X \coprod Y)}$.

Now let $\CC$ be a category with finite products. $\CC$ 
\emph{has $\bot$-maps} if for each object $A$ there is a distinguished 
morphism $\bot_{A} : \termobject \rightarrow A$. We then define 
$\bot_{A,B}$ for all $A$, $B$ by
\begin{displaymath}
	\bot_{A,B} = 
	A \stackrel{\tarrow_{A}}{\longrightarrow}  \termobject 
	\stackrel{\bot_{B}}{\longrightarrow}  B .
\end{displaymath}
We require that $\bot_{B,C} \circ f = \bot_{A,C}$ for all objects $A$, 
$B$, $C$ and $f : A \rightarrow B$; and that 
\[ \bot_{A \times B} = \langle \bot_{A} , \bot_{B} 
\rangle . \]

A morphism $f : A \rightarrow B$ is \emph{strict} if $f \circ 
\bot_{A} = \bot_{B}$. The strict morphisms form a sub-category which 
we denote $\CCs$. Note that $\CCs$ is enriched over $\Setst$. Thus 
there is an enriched hom-functor 
\begin{displaymath}
	\CCshom{-}{-} : \CCs^{\mathsf{o}} \times \CCs \longrightarrow \Setst .
\end{displaymath}
Note also that for any object $A$, 
\[ \CChom{A}{-} : \CCs \longrightarrow \Setst \]
is a well-defined functor, since the basepoint 
of $\CChom{A}{B}$ is $\bot_{A,B}$, and for any strict $f : B \rightarrow 
C$,
\begin{displaymath}
	f \circ \bot_{A,B} = \bot_{A,C} .
\end{displaymath}

A morphism $f : A \rightarrow B$ is \emph{total} if it is strict and 
$f \not= \bot_{A,B}$. We write $\CCt{A}{B}$ for the set of total 
morphisms from $A$ to $B$, so that
\begin{displaymath}
	\CCshom{A}{B} = \bp{\CCt{A}{B}} .
\end{displaymath}
However, note that total morphisms need not be closed under composition.

\paragraph{Examples}
In the category $\Cpo$ of directed-complete partial orders with least 
elements and continuous maps, strictness has its expected meaning.

In the categories of games in \cite{AJM,HO,M96b}, the $\bot$-maps are 
the empty strategies. The strict strategies $\sigma : A 
\rightarrow B$ are those which respond to the opening move by 
Opponent (which must be in $B$) with a move in $A$ if they have any 
response at all.

\subsection{Atomic and discrete objects}
Let $\CC$ be a category with finite products and $\bot$-maps.
An object $B$ of $\CC$ is a \emph{$\pi$-atom} (\textit{cf.} 
\cite{Joy92, Joy96}) if
\begin{displaymath}
	\CCshom{-}{B} : \CCs^{\mathsf{o}} \longrightarrow \Setst	
\end{displaymath}
preserves coproducts, \textit{i.e.} for each finite family $\{ A_{i} 
\mid i \in I \}$ of objects in $\CC$, the canonical map
\begin{displaymath}
	\sum_{i \in I} \CCshom{A_{i}}{B} = \bp{(\coprod_{i \in I} 
	\CCt{A_{i}}{B})} \longrightarrow \CCshom{\prod_{i \in I} A_{i}}{B}
\end{displaymath}
\begin{displaymath}
	\left\{ 
	\begin{array}{ccc}
		(i, f) & \mapsto & \pi_{i} ; f  \\
		\ast & \mapsto  & \bot_{\prod_{i \in I} A_{i} , B}
	\end{array}
	\right.
\end{displaymath}
is a bijection. The motivation for this terminology comes from lattice 
theory (Joyal is generalizing Whitman's theorem on free lattices). 
A $\pi$-atom in a lattice (also often called a meet-irreducible element, 
\textit{cf.} \cite{DP90}) is an element $a$ such that
\[ \bigwedge_{i=1}^{n} a_{i} \leq a \;\; \Longrightarrow \;\; \exists 
i . \, a_{i} \leq a. \]
Generalizing from posets to (enriched) categories, we get the definition given 
above.

An object $B$ is \emph{discrete} if for each $A$ the canonical map
\begin{displaymath}
	\CCshom{A}{B} + \CChom{\termobject}{B} \longrightarrow \CChom{A}{B}
\end{displaymath}
\begin{displaymath}
	f : A \rightarrow_{s} B \mapsto f : A \rightarrow B, \quad x : 
	\termobject \rightarrow B \mapsto A \rightarrow \termobject 
	\stackrel{x}{\rightarrow} B
\end{displaymath}
is a bijection.

The idea behind this definition is that any morphism into a discrete 
object is either strict or constant. It should be recalled that the 
coproduct on the left is the coalesced sum in $\Setst$; this allows 
the constant--$\bot$ morphism (which is both strict \emph{and} 
constant) to be properly accounted for.

We write $\CCpd$ for the full subcategory of $\CC$ determined by the 
$\pi$-atomic objects.

\paragraph{Examples}
In the (Linear) categories of games in \cite{AJM,HO,M96b}, any game $B$ with a 
unique initial question is $\pi$-atomic. The response to this unique 
initial question in $B$ made by a total strategy $\prod_{i 
\in I} A_{i} \rightarrow B$ must be in one of the 
games $A_{i}$. Making such a move  entails projecting the product 
onto the 
$i$'th factor.  Flat games---\textit{i.e.} 
those with a unique initial question and a set of possible answers, 
after which nothing more can happen---are discrete. This just says 
that if $A$ is discrete, then any strategy $A \rightarrow B$ is either 
the empty strategy, or responds to the unique initial question in $B$ 
with some move in $A$---and hence is strict; or responds with an 
answer in $B$ which completes the play, and hence is a ``constant'' strategy.

In $\Cpo$, flat domains are discrete (any continuous function into a 
flat domain is either strict or constant); $\Coh$, the category of 
coherence spaces and linear maps, is \emph{soft} in the sense of 
\cite{Joy92,Joy96}---see \cite{HJ97}.
\subsection{Standard datatypes}
Let $\CC$ be a category with $\bot$-maps as in Section~2.2. 
We assume given a class of objects of $\CC$ which we will call 
``well-opened'',
which forms an exponential ideal, \textit{i.e.} if $B$ is well-opened
os is $A \linimpl B$, and which moreover is closed under products.
We write $\CCwo$ for the sub-category of well-opened objects.

We say 
that $\CC$ \emph{has standard datatypes} if:
\begin{itemize}
\item The total maps on $\CCwo$ form a sub-category $\CCtwo$.

\item The functor
\begin{displaymath}
	\CChom{\termobject}{-} : \CCtwo \longrightarrow \Setst
\end{displaymath}
has a left adjoint left inverse $\widetilde{(\cdot )}$.
\end{itemize}

Unpacking this definition, for each well-opened object $A$ of $\CC$ 
and pointed set $\bp{X}$, 
there is an isomorphism
\begin{displaymath}
	\CCtwohom{\widetilde{\bp{X}}}{A} \cong \Setst (\bp{X}, 
	\CChom{\termobject}{A})
\end{displaymath}
natural in $\bp{X}$ and $A$, such that the unit
\begin{displaymath}
	\eta_{\bp{X}} : \bp{X} \longrightarrow \CChom{\termobject}{\widetilde{\bp{X}}}
\end{displaymath}
is an isomorphism. In particular, $\bp{X}$ is well-opened, and there are maps
\begin{displaymath}
	\bar{x} : \termobject \rightarrow \widetilde{\bp{X}} \quad (x \in X)
\end{displaymath}
with $\bar{x} \not= \bot_{\bpXst}$, and for each family
\begin{displaymath}
	(f_{x} : \termobject \rightarrow A \mid x \in X )
\end{displaymath}
a unique total morphism
\begin{displaymath}
	[ f_{x} \mid x \in X ] : \bpXst \rightarrow A
\end{displaymath}
such that
\begin{displaymath}
	\bar{x}_{0} ; [ f_{x} \mid x \in X ] = f_{x_{0}} \quad (x_{0} \in X).
\end{displaymath}
\paragraph{Examples}
In the categories of games in \cite{AJM,HO,M96b}, $\bpXst$ is the ``flat game'' 
with a unique initial question, and one answer for each $x \in X$.
The well-opened objects in these categories are those in which any move
that \emph{can} occur as the first move in a play can \emph{only} so occur.

In $\Cpo$, a slightly different situation prevails. The functor
\[ \Cpo (\termobject , -) : \Cpo_{s} \longrightarrow \Setst \]
has a left adjoint left inverse, which sends $\bp{X}$ to the flat 
domain $X_{\bot}$.

\section{Sequential Categories}

Let $\CC$ be an affine category with $\bot$-maps and standard 
datatypes. $\CC$ is a \emph{sequential category} if it satisfies the 
following axioms (A1)--(A5).

\paragraph{(A1)}
$\bpXst$ is discrete for each set $X$.

\paragraph{(A2)}
$\bpXst$ is $\pi$-atomic for each set $X$, and $\CCpd$ is an exponential ideal.

\paragraph{(A3)}
\textbf{(Uniformity of threads)}. Let $A$ and $B$ be well-opened
objects. Then
\begin{displaymath}
	\CChom{\llbang A}{\llbang B} \cong \CChom{\llbang A}{B} .
\end{displaymath}
More precisely, there are canonical maps
\begin{displaymath}
	\begin{array}{ccccc}
		f : \llbang A \rightarrow \llbang B & \mapsto  &  f ; \der_{B} 
		& : & \llbang A \rightarrow B  \\
		g : \llbang A \rightarrow B & \mapsto & g^{\dagger} 
		& : & \llbang A \rightarrow \llbang B
	\end{array}
\end{displaymath}
and since $({\llbang }, \der , (\cdot )^{{\dagger}})$ is a comonad,
\begin{displaymath}
	g^{\dagger} ; \der_{B} = g.
\end{displaymath}
The content of (A3) is that
\begin{displaymath}
	(f ; \der_{B})^{{\dagger}} = f
\end{displaymath}
\textit{i.e.} that the two passages are mutually inverse.

This property was proved for categories of games in \cite{AJM} and 
subsequently in \cite{M96b}, under the name of the ``Bang Lemma''. The 
idea is that morphisms $ f : \llbang A \rightarrow \llbang B$ must display uniform 
behaviour in all ``threads'', \textit{i.e.} in each copy of $B$ 
together with its associated copies of $A$. This property holds in 
these categories as a consequence of \emph{history-freeness} in 
\cite{AJM}, and of \emph{innocence} in \cite{M96b}. The idea in each 
case is that a strategy $ f : \llbang A \rightarrow \llbang B$ can only 
``see'' the current thread, and hence must play like the promotion of 
its restriction to a single thread, i.e. like $(f ; 
\der_{B})^{{\dagger}}$.

\paragraph{(A4)}
\textbf{(Linearization of Head Occurrence)}.
\begin{displaymath}
	\CCshom{\llbang A}{B} \cong \CCshom{A}{\llbang A \linimpl B} .
\end{displaymath}
More precisely, there is a canonical map
\begin{displaymath}
	\begin{array}{c}
	\CCshom{A}{\llbang A \linimpl B} \\
	\Big\downarrow\vcenter{%
	\rlap{$\Lambda^{-1}$}} \\
	\CChom{A \otimes \llbang A}{B} \\
	\Big\downarrow\vcenter{%
	\rlap{$\CChom{\der_{A} \otimes \ident_{\llbang A}}{\ident_{B}}$}} \\
	\CChom{\llbang A \otimes \llbang A}{B} \\
	\Big\downarrow\vcenter{%
	\rlap{$\CChom{\con_{A}}{\ident_{B}}$}} \\
	\CChom{\llbang A}{B}
	\end{array}
\end{displaymath}
The content of (A4) is firstly that this map factors through 
the inclusion
\begin{displaymath}
	\CCshom{\llbang A}{B} \hookrightarrow \CChom{\llbang A}{B}
\end{displaymath}
and secondly that the corestriction
\begin{displaymath}
	\CCshom{A}{\llbang A \linimpl B} \longrightarrow \CCshom{\llbang A}{B}
\end{displaymath}
is a bijection.

This property was proved, under the name of ``Separation of Head 
Occurrence'', for categories of games firstly in 
\cite{AJM}, and subsequently in \cite{M96b}.
There is a suggestive analogy, at least, with operational and 
proof-theoretic ideas of treating head variables linearly 
\cite{Gir87,DHR96}.
The idea is simply that we can split the use of many copies of $A$ 
in $\llbang A$ on the left into the tensor product of the 
\emph{first} copy used, of type $A$, and the remaining copies, of 
type $\llbang A$. In the case of strict strategies, which are either 
empty or use at least one copy of $A$, this correspondence is biunique.

\paragraph{(A5)}
\textbf{(Linear Function Extensionality)}.
There is an isomorphism
\begin{displaymath}
	\CCshom{A \linimpl B}{\llbang C \linimpl D} \cong \bp{\CChom{\llbang C}{A}} \otimes 
	\CCshom{B}{\llbang C \linimpl D}
\end{displaymath}
provided $B$ and $D$ are discrete. (The tensor product on the right is 
smash product in $\Setst$. Note that
\begin{displaymath}
	\bp{\CChom{\llbang C}{A}} \otimes \CCshom{B}{\llbang C \linimpl D} = 
	\bp{(\CChom{\llbang C}{A} \times \CCt{B}{\llbang C \linimpl D})} . )
\end{displaymath}
More precisely, there is a canonical map
\begin{displaymath}
	\begin{array}{c}
	\CChom{\llbang C}{A} \times \CCt{B}{\llbang C \linimpl D} \\
	\Big\downarrow\vcenter{%
	\rlap{$(-) \linimpl (-)$}} \\
	\CChom{A \linimpl B}{\llbang C \linimpl ( \llbang C \linimpl D)} \\
	\Big\downarrow\vcenter{%
	\rlap{$\cong$}} \\
	\CChom{A \linimpl B}{\llbang C \otimes \llbang C \linimpl D} \\
	\Big\downarrow\vcenter{%
	\rlap{$\CChom{\ident}{\con_{C} \linimpl \ident}$}} \\
	\CChom{A \linimpl B}{\llbang C \linimpl D}
	\end{array}	
\end{displaymath}
The content of (A5) is that this map factors through the 
inclusion
\begin{displaymath}
\CCt{A \linimpl B}{\llbang C \linimpl D} \hookrightarrow \CChom{A \linimpl 
B}{\llbang C \linimpl D}	
\end{displaymath}
and that the corestriction
\begin{displaymath}
	\CChom{\llbang C}{A} \times \CCt{B}{\llbang C \linimpl D} \longrightarrow 
	\CCt{A \linimpl 
B}{\llbang C \linimpl D}	
\end{displaymath}
and hence also the ``lifted'' map
\begin{displaymath}
	\bp{\CChom{\llbang C}{A}} \otimes 
	\CCshom{B}{\llbang C \linimpl D} \longrightarrow \CCshom{A \linimpl B}{\llbang C \linimpl D}
\end{displaymath}
is a bijection.

A special case of this property, under the same name, was proved for 
categories of games in \cite{AJM}, and subsequently in \cite{M96b}. The 
general case was implicit in the proof of the Decomposition Theorem 
in \cite{AJM}.
Intuitively, this axiom says that the only thing we can do with a 
linear functional parameter is to apply it to an argument, and apply 
some function to the result. The verification of this axiom in the 
categories of games considered in \cite{AJM,M96b} makes essential 
use \emph{both} of history-freeness/innocence, \emph{and} of well-bracketedness.
The idea is that a strict strategy of the type displayed in the axiom 
must respond to the initial question in $D$ by ``calling its function 
argument'', i.e. by making the (unique) initital move in $B$. By the 
bracketing condition, the initial question in $D$ cannot be answered 
until the initial question in $B$ has been answered, i.e. until the 
play in $B$ is complete. This allows us to decompose the strategy we 
started with into sub-strategies, corresponding to what is done before 
and after the play in $B$ is completed. History-freeness/innocence 
then guarantess that the continuation strategy which proceeds after 
the play in $B$ is completed can depend \emph{only} on the answer 
returned, not on the interaction which took place between the 
function and its arguments. This is one of the key points where 
essentially non-functional behaviour is being excluded.

\paragraph{Examples}
A minor embarassment is that neither of our two concrete examples of 
models for the above axioms, namely the categories of games described 
in \cite{AJM,M96b}, \emph{quite} succeeds in being a sequential 
category! They \emph{do} satisfy the key axioms (A1)--(A5). In the case of the 
category in \cite{AJM}, the problem is that it fails to have products in the 
underlying 
Affine category---although the co-Kleisli category \emph{is} cartesian closed. 
However, there is a ``candidate'' for the product in the Affine 
category---which turns into a real product in the co-Kleisli 
category---which \emph{does} have projections, and with respect to which the 
required properties relating to $\pi$-atomicity do hold. This is 
enough for our applications to definability to follow.	Similarly, in the 
category used in \cite{M96b} $!$ fails to be a co-monad; however, one 
does get a cartesian closed co-Kleisli category by restricting to the 
well-opened objects.

These minor mismatches are probably best taken as an indication that 
we do not yet have a definitive formulation of game semantics.
\section{Decomposition}
We will now prove a decomposition theorem in the sense of \cite{AJM} from 
the axioms.

Let $\CC$ be a sequential category. Let
\begin{displaymath}
	A_{i} = \llbang (B_{i,1} \times \cdots \times B_{i, q_{i}}) \linimpl 
	\bpXst \quad (1 \leq i \leq k)
\end{displaymath}
be objects of $\CC$. We write $\vec{A} = A_{1} \times \cdots \times 
A_{k}$. Consider a morphism $f : \llbang \vec{A} \rightarrow \bpXst$. By 
(A1), $\bpXst$ is discrete, hence three disjoint cases apply:
\begin{itemize}
	\item  $f = \bot$.

	\item  $f = \llbang \vec{A} \rightarrow \termobject 
	\stackrel{g}{\rightarrow} \bpXst$.
	In this case, by the universal property of $\bpXst$, we must have 
	$g = \bar{x}$ for some unique $x \in X$.

	\item  $f$ is total. In this case, by (A4) $f$ is in 
	bijective correspondence with a total morphism
	\begin{displaymath}
		f' : \vec{A} \rightarrow \llbang \vec{A} \linimpl \bpXst .
	\end{displaymath}
	By (A2), $f'$ factors by a projection $\pi_{i}$ through
	\begin{displaymath}
		f_{i} : A_{i} \rightarrow_{t} \llbang \vec{A} \linimpl \bpXst
	\end{displaymath}
	for a unique $i$, $1 \leq i \leq k$. By (A5), $f_{i}$ 
	decomposes into
	\begin{displaymath}
		g : \llbang \vec{A} \rightarrow \llbang \vec{B}_{i} , \quad h : \bpXst 
		\rightarrow_{t} (\llbang \vec{A} \linimpl \bpXst )
	\end{displaymath}
	where $\vec{B}_{i} = B_{i,1} \times \cdots \times B_{i,q_{i}}$. By 
	(A3), $g = (g ; \der_{\vec{B}_{i}})^{\dagger}$, and by the 
	universal property of the product,
	\begin{displaymath}
		g = \langle g_{1}, \ldots , g_{q_{i}} \rangle^{\dagger}
	\end{displaymath}
	for unique $g_{j} : \llbang \vec{A} \rightarrow B_{i,j}$, $1 \leq j \leq q_{i}$. 
	By the universal property of $\bpXst$,
	\begin{displaymath}
		h = [ h_{x} : \termobject \rightarrow \llbang \vec{A} \linimpl \bpXst 
		\mid x \in X ] .
	\end{displaymath}
	Thus we obtain
	\begin{displaymath}
			f   =   \Comi{g_{1}, \ldots , g_{q_{i}}}{h_{x} \mid x \in X}  
	\end{displaymath}
	where $\Comi{g_{1}, \ldots , g_{q_{i}}}{h_{x} \mid x \in X}$ 
	abbreviates
	\begin{displaymath}
			   \con_{\vec{A}} ; (\con_{\vec{A}} ; 
			   (\der_{\vec{A}} ; \pi_{i}) \otimes 
			   \langle g_{1}, \ldots , g_{q_{i}} \rangle^{\dagger} ; \Ap ) \otimes 
			   \ident_{\llbang \vec{A}} ; \\
			   \Lambda^{-1}([ h_{x} \mid x \in X]) .
	\end{displaymath}
	\end{itemize}
	We summarize this analysis in
	\begin{theorem}[Decomposition]
	With notation as above, one of the following three cases applies:
	\begin{itemize}
	\item  $f = \bot$.

	\item  $f = \llbang \vec{A} \rightarrow \termobject 
	\stackrel{\bar{x}}{\rightarrow} \bpXst$.
	
	\item $f = \Comi{g_{1}, \ldots , g_{q_{i}}}{h_{x} \mid x \in X}$.
	\end{itemize}
	Moreover, this decomposition is unique.
	\end{theorem}

\section{PCF}
In this section we briefly recall the language PCF, its operational 
semantics and observational preorder, in a streamlined version 
convenient for our purposes.

PCF is an applied simply typed $\lambda$-calculus with a single base 
type $\mathbf{nat}$. The constants are as follows:
\begin{itemize}
	\item  recursion combinators $\YY_{T} : (T \Rightarrow T) \Rightarrow 
	T$ for each type $T$.

	\item  $\Omega : \nat$.

	\item  $\underline{n} : \nat$ for each $n \in \Nat$.

	\item  $\casek : \nat \Rightarrow \underbrace{\nat \Rightarrow \cdots 
	\Rightarrow \nat}_{k} \Rightarrow \nat$ for each $k \in \Nat$.
\end{itemize}
The main difference from PCF as originally presented by Scott and 
Plotkin is in the use of the $\casek$ constants instead of the more 
familiar conditionals and arithmetic operations. The $\casek$ 
constants are needed for a precise correspondence at the intensional 
level, and as shown in detail in \cite{AJM}, the difference is 
insignificant as far as observational equivalence is concerned.

The operational semantics is defined via a structural congruence 
$\equiv$ (\textit{cf.} \cite{Mil92}) and an evaluation relation 
$\underline{\ } \Converges \underline{\ }$. The structural congruence 
is the congruence on terms generated by $\beta \eta$-conversion and 
all instances of
 \begin{displaymath}
 	\YY M \equiv M(\YY M) .
 \end{displaymath}
 The evaluation relation $P \Converges \underline{n}$ is defined 
 between \emph{programs}, \textit{i.e.} closed terms of type $\nat$, 
 and numerals $\underline{n}$, inductively as follows:
 \begin{displaymath}
 	\frac{M \equiv M' \quad M' \Converges \underline{n}}{M \Converges 
 	\underline{n}}
 	\quad \quad
 	\frac{}{\underline{n} \Converges \underline{n}}
 \end{displaymath}
 \begin{displaymath}
 	\frac{P \Converges \underline{i} \;\; (i < k) \quad P_{i} 
 	\Converges \underline{n}}{\casek P P_{0} \cdots P_{k-1} \Converges 
 	\underline{n}}
 \end{displaymath}
 Let $\Trm{\Gamma}{T}$ be the set of terms $M$ such that $\Gamma \vdash 
 M : T$ is derivable. Let $\Ctxt{\Gamma}{T}$ be the set of contexts 
 $C[\cdot ]$ such that $C[M]$ is a program for all $M \in 
 \Trm{\Gamma}{T}$. The observational preorder is defined at $\Gamma , 
 T$ by:
 \begin{displaymath}
 	M \preord_{\Gamma , T} N \;\; \Longleftrightarrow \;\; \forall 
 	C[\cdot ] \in \Ctxt{\Gamma}{T} . \, C[M] \Converges \underline{n} \;
 	\Longrightarrow \; C[N] \Converges \underline{n} .
 \end{displaymath}
 \section{Computational Adequacy}
 Let $\CC$ be an affine category with $\bot$-maps and standard 
 datatypes. The cartesian closed category $K_{\llbang }(\CC )$ provides 
 a model of the fragment of PCF obtained by omitting the recursion 
 combinators $\YY_{T}$. The base type $\nat$ is interpreted by 
 $\Natst$, the constants $\underline{n}$ by $\overline{n}$, $n \in 
 \Nat$, and $\Omega$ by $\bot_{\Natst}$. The constant $\casek$ is 
 interpreted by
 \begin{displaymath}
  \der_{\Natst} ; [ f_{i} \mid i \in \Nat ]
 \end{displaymath}
 where
 \begin{displaymath}
 	f_{i} = \left\{ 
 	\begin{array}{ll}
 		\Lambda^{k} (\pi_{i}) , & 0 \leq i < k  \\
 		\bot ,  & i \geq k .
 	\end{array}
 	\right.
 \end{displaymath}
 This interpretation is extended to all terms in the standard way 
 \cite{Cro94}.
 
 To accommodate recursion, we need another definition. Let $\KK$ be a 
 cartesian closed category. A \emph{fixpoint 
 operator} on $\KK$ is a family of maps
 \begin{displaymath}
 	\fix{(\ )_{A}} : \KK (A, A) \longrightarrow \KK (\termobject , A)
 \end{displaymath}
 satisfying
 \begin{displaymath}
 	f \circ \fix{f} = \fix{f} .
 	\label{fix1}
 \end{displaymath}

 Given such an operator, we can interpret the fixpoint combinator 
 $\YY_{A} : \termobject \rightarrow (A \Rightarrow A) \Rightarrow A$ 
 by
 \begin{displaymath}
 	\fix{\lsem F : (A \Rightarrow A) \Rightarrow A \vdash \lambda f : A 
 	\Rightarrow A . \, f (Ff) \rsem} .
 \end{displaymath}
 A model is said to be \emph{computationally adequate} if, for all 
 programs $P$ and $n \in \Nat$:
 \begin{displaymath}
 	P \Converges \underline{n} \; \Longleftrightarrow \; \lsem P 
 	\rsem = \overline{n}.
 \end{displaymath}
 Let $\CC$ be an affine category with $\bot$-maps and standard 
 datatypes, equipped with a fixpoint operator on $K_{\llbang }(\CC )$.
 $\CC$ is said to be \emph{continuously observable} if, for all $f : 
 A \rightarrow A$ and $g : A \rightarrow \bpXst$ in $K_{\llbang }(\CC )$, 
 and all $x \in X$:
 \begin{displaymath}
 	g \circ \fix{f} = \overline{x} \; \Longleftrightarrow \; \exists k 
 	\in \omega . \, g \circ f^{k} \circ \bot = \overline{x} .
 \end{displaymath}
Recall from \cite{AJM} that a \emph{rational cartesian closed 
category} is a cartesian closed category enriched over pointed 
posets, with least upper bounds of ``definable'' chains, i.e. 
those of the form $(f^{k} \circ \bot \mid k \in \omega )$. This 
provides just enough structure to interpret the recursion 
combinators $\YY$ in 
terms of least fixpoints.
It is shown in \cite{AJM} that the category of games studied there is 
rational; this, together with the fact that the ``points'' of the standard 
datatypes 
form flat domains implies that the category is continuously observable.
 \begin{theorem}[Computational Adequacy]
 If $\CC$ is continuously observable then $K_{\llbang }(\CC )$ is a 
 computationally adequate model of PCF.
\end{theorem}
The original proof by Plotkin for the Scott continuous function model 
\cite{Plo77} goes through in our axiomatic setting (\textit{cf.} 
\cite{Brau96}).

In practice, it is often more convenient to verify somewhat stronger 
axioms. We say that a sequential category $\CC$ is \emph{normed} if:
\begin{itemize}
	\item  It is enriched over algebraic cpo's, with the $\bot$-morphisms 
	being the least elements in the partial orderings, and such that 
	$\CChom{\termobject}{\bpXst}$ is order-isomorphic to the flat domain 
	$X_{\bot}$.

	\item  There is a norm function $\norm{f} \in \Nat$ on \emph{compact} 
	morphisms $f$, such that, for each PCF type $T$ and compact $f : 
	\termobject \longrightarrow \lsem T \rsem$:
		\[ f = \Comi{f_{1} , \ldots , f_{q_{i}}}{g_{n} \mid n \in \Nat} \;\; 
		\Longrightarrow \] 
 	\[ \sup (\{  \norm{f_{j}}  \mid 1 \leq j \leq q_{i} \} \cup \{ 
 	\norm{g_{n}} \mid n \in \Nat \} ) < \norm{f} , \]
 	and $g_{n} = \bot$ for almost all $n \in \Nat$.
\end{itemize}
Note that a normed category is automatically observably continuous and 
hence computationally adequate.

The categories in \cite{HO,M96b} are normed, with the norm of a compact 
innocent strategy given by the cardinality of its view function.
 \section{Definability}
 Let $T = T_{1} \Rightarrow \cdots T_{k} \Rightarrow \nat$ be a PCF 
 type. Note that
 \begin{displaymath}
 	\begin{array}{lcl}
 	\lsem T \rsem &=& \lsem T_{1} \rsem \Rightarrow \cdots \lsem T_{k} \rsem 
 	\Rightarrow \lsem \nat \rsem \\
 	&=& \llbang \lsem T_{1} \rsem \linimpl \cdots \llbang \lsem T_{k} \rsem \linimpl 
 	\Natst \\
 	&\cong& (\llbang \lsem T_{1} \rsem \otimes \cdots \otimes \llbang \lsem T_{k} 
 	\rsem ) \linimpl \Natst \\
 	&\cong& \llbang (\lsem T_{1} \rsem \times \cdots \times \lsem T_{k} \rsem ) 
 	\linimpl \Natst . \\
 	\end{array}
 \end{displaymath}
 To save notational overhead, we shall elide this canonical 
 isomorphism, equating the ``curried'' and ``uncurried'' versions of 
 types. (However, for a careful treatment in which these isomorphisms 
 are made explicit, see \cite{AJM}).
 
 Let $\CC$ be a sequential category. For each $f : \termobject 
 \rightarrow \lsem T \rsem$ in $\CC$ and $k \in \omega$, we define 
 $p_{k}(f)$ inductively as follows:
 \[ \begin{array}{lcl}
 p_{0}(f) & = & \bot \\
 p_{k+1} (f) & = & \left\{ \begin{array}{ll}
 \bot , & f = \bot \\
 f, & f = \tarrow ; \overline{n} \\
 f' , & f = \Comi{f_{1},\ldots , f_{q_{i}}}{g_{n} \mid n \in \Nat}
 \end{array}
 \right. 
 \end{array} \]
 where
 \[ \begin{array}{lcl}
 f' & = & \Comi{p_{k}(f_{i}), \ldots . p_{k}(f_{q_{i}})}{h_{n} \mid n 
 \in \Nat} \\
 h_{n} & = & \left\{ \begin{array}{ll}
 p_{k}(g_{n}) , & 0 \leq n < k \\
 \bot & n \geq k .
 \end{array}
 \right.
 \end{array} \]
 This definition by cases is valid by the Decomposition Theorem.
 
 \begin{theorem}[Definability]
 For each PCF type $T$, $f : \termobject \rightarrow \lsem T \rsem$ 
 and $k \in \omega$, $p_{k}(f)$ is definable in PCF. That is, there 
 exists a PCF term $\vdash M : T$ such that $p_{k}(f) = \lsem M 
 \rsem$.
 \end{theorem}
 
 \proof\  By induction on $k$, and cases on the decomposition of $f$. We 
 write $T = \widetilde{T} \Rightarrow \nat$, where $\widetilde{T} = T_{1}, 
 \ldots , T_{k}$.
 \[ \begin{array}{lcccl}
 p_{0}(f) & = & \bot & = & \lsem \lambda \tilde{x} : \widetilde{T}. \, 
 \Omega \rsem \\
 p_{k+1}(\bot) & = & \bot & = & \lsem \lambda \tilde{x} : \widetilde{T}. \, 
 \Omega \rsem \\ 
 p_{k+1}(\tarrow ; \overline{n}) & = & \tarrow ; \overline{n}  & = & \lsem 
 \lambda 
 \tilde{x} : \widetilde{T}. \, \underline{n} \rsem 
 \end{array} \]
\[ p_{k+1}(\Comi{p_{k}(f_{i}), \ldots . p_{k}(f_{q_{i}})}{h_{n} \mid n 
 \in \Nat}) = \]
 \[ \lsem \lambda \tilde{x} : \widetilde{T} . \, \casek (x_{i} 
 (M_{1}\tilde{x}) \cdots (M_{q_{i}}\tilde{x})) (P_{0}\tilde{x}) \cdots 
 (P_{k-1}\tilde{x}) \rsem  \]
 where
\[ \begin{array}{lcl}
 p_{k}(f_{j}) & = & \lsem M_{j} \rsem , \quad 1 \leq j \leq q_{i}, \\
 p_{k}(g_{n}) & = & \lsem P_{n} \rsem , \quad 0 \leq n < k. \quad \qed
 \end{array} \]
 For normed categories, we can prove a stronger result.
 \begin{theorem}[Definability for Normed Sequential Categories]
 For each PCF type $T$ and compact $f : \termobject \rightarrow \lsem T \rsem$, 
 $f$ is definable in PCF. That is, there 
 exists a PCF term $\vdash M : T$ such that $f = \lsem M 
 \rsem$.
 \end{theorem}
 
 \proof\  By complete induction on $\norm{f}$, and cases on the decomposition 
 of $f$. We 
 write $T = \widetilde{T} \Rightarrow \nat$, where $\widetilde{T} = T_{1}, 
 \ldots , T_{k}$.
 If $f = \bot$, then
 \[ f  = \lsem \lambda \tilde{x} : \widetilde{T}. \, 
 \Omega \rsem . \]
 If $f = \tarrow ; \overline{n}$, then  
 \[ f = \lsem 
 \lambda 
 \tilde{x} : \widetilde{T}. \, \underline{n} \rsem . \]
 If $f = \Comi{f_{1} , \ldots , f_{q_{i}}}{g_{n} \mid n \in \Nat}$, 
 then
 \[ f = \lsem \lambda \tilde{x} : \widetilde{T} . \, \casel (x_{i} 
 (M_{1}\tilde{x}) \cdots (M_{q_{i}}\tilde{x})) (P_{0}\tilde{x}) \cdots 
 (P_{k-1}\tilde{x}) \rsem  \]
 where by induction hypothesis
\[ \begin{array}{lcl}
 f_{j} & = & \lsem M_{j} \rsem , \quad 1 \leq j \leq q_{i}, \\
 g_{n} & = & \lsem P_{n} \rsem , \quad 0 \leq n < l ,
 \end{array} \]
and $g_{m} = \bot$ for all $m \geq l$. $\quad \qed$

We note that the terms used to exhibit definability in Theorems~7.1 and~7.2
are of a special form; the \emph{evaluation trees}
of \cite{AJM}.
These can be seen as a form of Bohm tree appropriate for PCF.
These trees were in fact first identified as the right notion of normal
forms for PCF terms as a consequence of the work on game semantics; this
is an interesting example of feedback from semantics to syntax.
In \cite{AJM} it is shown that there is an order-isomorphism between
the (possibly infinite) evaluation trees at any PCF type, and the
strategies for the game denoted by that type. 
A similar result
can be obtained at the axiomatic level of this paper. Since it is not
needed in order to obtain the full abstraction results, we 
will omit the details.

 \section{Full Abstraction}
 Let $\CC$ be a poset-enriched model of PCF. $\CC$ is \emph{sound} 
 if, for all $M, N \in \Trm{\Gamma}{T}$,
 \begin{displaymath}
 	\lsem \Gamma \vdash M : T \rsem \leqslant \lsem \Gamma \vdash N : T 
 	\rsem \; \Longrightarrow \; M \sqpreord_{\Gamma , T} N ,
 \end{displaymath}
 and \emph{complete} if the converse holds. $\CC$ is \emph{fully 
 abstract} if it is both sound and complete.
 \begin{lemma}
 $\CC$ is fully abstract iff soundness and completeness hold for all 
 \emph{closed} terms.
 \end{lemma}
 
 \proof\  The left-to-right implication is immediate. For the 
 converse, we show firstly that, if $\Gamma = x_{1} : T_{k}, \ldots , x_{k} : 
 T_{k}$,
 \begin{equation}
 	M \sqpreord_{\Gamma , T} N \; \Longleftrightarrow \; \lambda 
 	\tilde{x} : \tilde{T} . \, M \sqpreord_{\varnothing , T} \lambda 
 	\tilde{x} : \tilde{T} . \, N.
 	\label{fal1}
 \end{equation}
 Again, the left-to-right implication is immediate. For the converse, 
 assume that $\lambda 
 	\tilde{x} : \tilde{T} . \, M \sqpreord_{\Gamma , T} \lambda 
 	\tilde{x} : \tilde{T} . \, N$ and that $C[M] \Converges 
 	\underline{n}$. Define a new context
 	\begin{displaymath}
 		D[\cdot ] = C[[\cdot ]\tilde{x}] .
 	\end{displaymath}
 	Then $D[\lambda \tilde{x} : \tilde{T} . \, M] \equiv C[M]$ and hence 
 	$D[\lambda \tilde{x} : \tilde{T} . \, M] \Converges \underline{n}$. 
 	This implies that 
 	$D[\lambda \tilde{x} : \tilde{T} . \, N] \Converges \underline{n}$, 
 	but since $D[\lambda \tilde{x} : \tilde{T} . \, N] \equiv C[N]$, we 
 	conclude that $C[N] \Converges \underline{n}$, as required.
 	
 	Furthermore since currying $\Lambda_{A,B,C} : \CC (A \times B, C) 
 	\longrightarrow \CC (A, B \Rightarrow C)$ is an order-isomorphism, it is 
 	immediate  that
 	\begin{equation}
 		\lsem \Gamma \vdash M : T \rsem \leqslant \lsem \Gamma \vdash N : T 
 		\rsem \; \Longleftrightarrow \; \lsem \vdash \lambda \tilde{x} : 
 		\tilde{T} . \, M \rsem \leqslant \lsem  \vdash \lambda \tilde{x} : 
 		\tilde{T} . \, N \rsem .
 		\label{fal2}
 	\end{equation}
 	By assumption,
 	\begin{equation}
 		\lambda \tilde{x} : 
 		\tilde{T} . \, M  \sqpreord_{\tilde{T} \Rightarrow T} 
 		\lambda \tilde{x} : 
 		\tilde{T} . \, N
 		\; \Longleftrightarrow \;
 		\lsem \lambda \tilde{x} : 
 		\tilde{T} . \, M \rsem \leqslant
 		\lsem \lambda \tilde{x} : 
 		\tilde{T} . \, N \rsem .
 		\label{fal3}
 	\end{equation}
 	Combining (\ref{fal1}), (\ref{fal2}) and (\ref{fal3}) we conclude 
 	that $\CC$ is fully abstract. $\quad \qed$
 	
 	Now let $\CC$ be an observably continuous sequential category. 
 	$K_{\llbang }(\CC )$ is a computationally adequate model of PCF, by the 
 	Computational Adequacy Theorem. We define the \emph{intrinsic 
 	preorder} $\ipreord_{A}$ on each $\KCC (\termobject , A)$ by 
 	\begin{displaymath}
 		f \ipreord_{A} g  \; \Longleftrightarrow \; \forall \alpha : A 
 		\rightarrow \Natst . \, \alpha \circ f = \overline{n} \; 
 		\Rightarrow \; \alpha \circ g = \overline{n} .
 	\end{displaymath}
 	This is extended to general homsets $\KCC (A, B)$ via the names of 
 	morphisms:
 	\begin{displaymath}
 		f \ipreord_{A, B} g \; \Longleftrightarrow \; \name{f} \ipreord_{A 
 		\Rightarrow B} \name{g}.
 	\end{displaymath}
 
 	\begin{proposition}
 	For all $A$, $B$ $\ipreord_{A, B}$ is a preorder, with least 
 	element $\bot_{A, B}$, and $\KCC$ is an enriched cartesian closed 
 	category with respect to this preorder. The poset reflection 
 	$\KCC / {\ipreord}$ is a rational cartesian closed category, and 
 	there is a full cartesian closed functor $Q : \KCC \rightarrow 
 	\KCC / {\ipreord}$ with the evident universal property which 
 	translates the interpretation of PCF in $\KCC$ to that in $\KCC / 
 	{\ipreord}$.
 	\end{proposition}
 	
 	\proof\ See \cite{HO,M96b,Brau96}. $\quad \qed$
 	
 	We say that $\CC$ is \emph{approximating} if for all PCF types $T$, 
 	$f : \lsem T \rsem \rightarrow \bpXst$ and $\alpha : \termobject 
 	\rightarrow \lsem T \rsem$ in $\KCC$, and $x \in X$:
 	\begin{displaymath}
 		f \circ \alpha = \overline{x} \; \Longleftrightarrow \; \exists k 
 		\in \omega . \, p_{k}(f) \circ \alpha = \overline{x} .
 	\end{displaymath}
 	In \cite{AJM} it is proved that the category of games considered 
 	there is approximating in Section~3.4.
 	\begin{theorem}
 	Let $\CC$ be an approximating sequential category. Then $\KCC / 
 	{\ipreord}$ is a fully abstract model of PCF.
 	\end{theorem}
 	
 	\proof\ By the Lemma, it suffices to show soundness and
 	completeness for closed terms. Note that if $M$ is closed, $C[M] 
 	\equiv (\lambda x. \, C[x]) M$ ($x \not\in FV(M)$), and hence $C[M] 
 	\Converges \underline{n} \; \Longleftrightarrow \; D[M] \Converges
 	\underline{n}$, where $D[\cdot ] = (\lambda x. \, C[x])[\cdot ]$. 
 	Thus $T$-contexts reduce to applications of functions of type $T 
 	\Rightarrow \nat$.
 	
 	Suppose then that $\lsem M \rsem \leqslant \lsem N \rsem$, and that 
 	$C[M] \Converges \underline{n}$. By computational adequacy, $\lsem 
 	C[M] \rsem = \overline{n}$, \emph{i.e.} $f \circ \lsem M \rsem = 
 	\overline{n}$, where $f = \lsem \lambda x.\, C[x] \rsem$. This 
 	implies that $f \circ \lsem N \rsem = 
 	\overline{n}$ \emph{i.e.} $\lsem 
 	C[N] \rsem = \overline{n}$, and by computational adequacy again, 
 	$C[N] \Converges \underline{n}$. This establishes soundness.
 	
 	For completeness, suppose that $\lsem \vdash M : T \rsem 
 	\nleqslant 
 	\lsem \vdash N : T \rsem$, \emph{i.e.} for some $f : \lsem T \rsem 
 	\rightarrow \Natst$, $f \circ \lsem M \rsem = \overline{n}$ and 
 	$f \circ \lsem N \rsem \not= \overline{n}$. Since $\CC$ is 
 	approximating, for some $k \in \omega$, $p_{k}(f) \circ \lsem M 
 	\rsem = \overline{n}$ and $p_{k}(f) \circ \lsem N 
 	\rsem \not= \overline{n}$. By the Definability Theorem, for some 
 	$P$, $p_{k}(f) = \lsem P \rsem$ and hence, defining $C[\cdot ] = 
 	P[\cdot ]$, $\lsem C[M] \rsem = p_{k}(f) \circ \lsem M \rsem = 
 	\overline{n}$ while $\lsem C[N] \rsem = p_{k}(f) \circ \lsem N \rsem \not= 
 	\overline{n}$. By computational adequacy, $C[M] \Converges 
 	\underline{n}$ while $\neg (C[N] \Converges \underline{n})$, and 
 	hence $\neg (M \sqpreord_{T} N)$, as required.
 	$\quad \qed$
 	
 	We have the corresponding result for normed sequential categories.
    \begin{theorem}
 	Let $\CC$ be a normed sequential category. Then $\KCC / 
 	{\ipreord}$ is a fully abstract model of PCF.
 	\end{theorem}
 	
 	\proof\ The proof is almost identical to that of the preceding 
 	theorem. The only difference is in the argument for completeness. 
 	Given the separating morphism $f : \lsem T \rsem 
 	\rightarrow \Natst$, we use the compactness of $\overline{n}$ in 
 	$\CChom{\termobject}{\Natst}$, the algebraicity of $\CChom{\lsem T 
 	\rsem}{\Natst}$, and the continuity of composition to obtain a 
 	\emph{compact} $f_{0} : \lsem T \rsem 
 	\rightarrow \Natst$ such that $f_{0} \circ \lsem M \rsem = \overline{n}$ and 
 	$f_{0} \circ \lsem N \rsem \not= \overline{n}$. Then we use the 
 	Definability Theorem for normed sequential categories to obtain the 
 	separating context $C[\cdot ]$. $\quad \qed$
 	
 \section{Full Completeness}
 Just as PCF is prototypical for higher-order programming languages, 
 so is the pure simply typed $\lambda$-calculus for logical systems. 
 (Definability results for game semantics of the pure calculus are 
 discussed in \cite{DHR96}, and were already known to the authors of 
 \cite{AJM,HO}.)
 
 We shall indicate how our axiomatic approach can be modified (in 
 fact: simplified) to deal with the pure calculus.
 
 We define a \emph{pure} sequential category to be an affine 
 category $\CC$ with a specified subcategory $\CCs$, with the same 
 objects as $\CC$, which has an 
 initial object $\iota$ (initial in $\CCs$), satisfying the following 
 axioms:
 
 \paragraph{(a1)} $\CChom{A}{\iota} = \CCshom{A}{\iota}$ for 
 all $A$.
 
 \paragraph{(a2)} $\iota$ is $\pi$-atomic (meaning that 
 $\CCshom{-}{\iota} : \CCs^{\mathsf{o}} \longrightarrow \Set$ preserves finite 
 coproducts), and the $\pi$-atomic objects form an exponential ideal.
 
 \paragraph{(a3)} \textbf{Uniformity of Threads}. As in Section~3.
 
 \paragraph{(a4)} \textbf{Linearization of Head Occurrence}. As in 
 Section~3.
 
 \paragraph{(a5)} \textbf{Linear Function Extensionality}.
 \begin{displaymath}
 	\CCshom{A \linimpl \iota}{\llbang C \linimpl \iota} \cong \CChom{\llbang C}{A} .
 \end{displaymath}
 More precisely, the canonical map
 \begin{displaymath}
 	(-) \linimpl \ident_{\iota} : \CChom{\llbang C}{A} \longrightarrow 
 	\CChom{A \linimpl 
 	\iota}{\llbang C \linimpl \iota} 
 \end{displaymath}
 is asserted to corestrict bijectively onto $\CCshom{A \linimpl 
 	\iota}{\llbang C \linimpl \iota}$. Note by the way that, since $\iota$ is 
 	initial in $\CCs$, $\CCshom{\iota}{\iota} = \{ \ident_{\iota} \}$.
 	
 	The following decomposition theorem can then be proved for any 
 	\[ f : \llbang (\prod_{i \in I} A_{i}) \longrightarrow \iota , \] 
 	where
 	\begin{displaymath}
 		A_{i} = \llbang (\prod_{j = 1}^{q_{i}} B_{i,j}) \linimpl \iota , \qquad 
 		1 \leq i \leq k .
 	\end{displaymath}
 	\begin{theorem}[Decomposition]
 	\begin{displaymath}
 		\begin{array}{lcl}
 		f &=& \Commi{g_{1}, \ldots g_{q_{i}}} \\
 		  &\stackrel{\Delta}{=}& \con_{\vec{A}} ; (\der_{\vec{A}} ; \pi_{i}) 
 		  \otimes 
 		  \langle g_{1}, \ldots , g_{q_{i}} \rangle^{\dagger} ; \Ap
 		\end{array}
 	\end{displaymath}
 	for a unique $i$, $1 \leq i \leq k$, and 
 	\[ f_{j} : \llbang (\prod_{i=1}^{k} 
 	A_{i}) \rightarrow B_{i,j}, \qquad 1 \leq j \leq q_{i} . \]
 	\end{theorem}
 	
 	\proof\ By (a1), $f$ is strict. By (a4) we 
 	obtain a strict morphism $f' : \vec{A} \rightarrow (\vec{A} 
 	\linimpl \iota )$, and by (a2) this factors by a unique 
 	$\pi_{i}$, $1 \leq i \leq k$, through
 	\[ f_{i} : A_{i} \rightarrow (\vec{A} \linimpl \iota ) . \]
 	By (a5) we obtain $g : \llbang  \vec{A} \rightarrow 
 	\vec{B}_{i}$, and by (a3) and the universal property of 
 	the product we obtain
 	\[ g = \langle g_{1}, \ldots , g_{q_{i}} \rangle^{\dagger} .
 	\quad \qed \]
 	
 	We replace the continuity postulates appropriate for PCF by a 
 	finiteness axiom. We stipulate that each morphism $f : A \rightarrow 
 	B$ in $\CC$ has a norm $\norm{f}_{A, B} \in \Nat$, such that
 	\[ f = \Commi{g_{1} , \ldots , g_{q_{i}}} \; \Rightarrow \; 
 	\sup_{j = 1, \ldots , q_{i}} \norm{g_{j}} < \norm{f} . \]
 	
 	\begin{theorem}[Definability]
 	Let $T$ be a simple type built from $\iota$. Then every $f : 
 	\termobject \rightarrow \lsem T \rsem$ in $\CC$ is definable, and in 
 	fact is the denotation of a unique long-$\beta\eta$-normal form.
 	\end{theorem}
 	
 	\proof\ By complete induction on $\norm{f}$. If $T = \widetilde{T} 
 	\Rightarrow \iota$ and $f = \Commi{g_{1} , \ldots , g_{q_{i}}}$, then
 	\[ f = \lsem \lambda \tilde{x} : \widetilde{T} . \, x_{i} 
 	(M_{1}\tilde{x}) \cdots (M_{q_{i}} \tilde{x}) \rsem \]
 	where $g_{j} = \lsem M_{j} \rsem$, $1 \leq j \leq q_{i}$. $\quad \qed$
 	
 	\paragraph{Examples}
 	The intended examples are the versions of \cite{AJM,HO,M96b} in which only 
 	total, compact strategies are included. Strictness of a strategy 
 	$\sigma : A \rightarrow B$ means that it responds to the opening 
 	move by moving in $A$. The interpretation of $\iota$ is as the game 
 	with a single move, which 
 	is an Opponent question.
 	
 	The above result is essentially a characterization of the free 
 	cartesian closed category generated by the one-object one-morphism 
 	category. More general characterization results can probably be 
 	developed along similar lines.
 	
 	Since under the Curry-Howard isomorphism \cite{GLT} the pure simply-typed 
 	$\lambda$-calculus corresponds to minimal implicational logic, this 
 	result has some relevance for Proof Theory. An interesting contrast 
 	with the full completeness results proved for Multiplicative Linear 
 	Logic in \cite{AJ94} and a number of other subsequent works is that 
 	various notions of ``uniformity'', dinaturality etc. play an 
 	important role in those results, but do not arise here.

 \paragraph{Universality} In \cite{AJM,HO} a stronger result than 
 Full Abstraction is proved, namely \emph{Universality}, i.e. that 
 \emph{all recursive strategies are definable in PCF}; or, 
 equivalently, that the model consisting of just the recursive 
 strategies has \emph{all} its elements definable. This is the 
 strongest possible definability result, and is closely related to the 
 notion of ``Logical Full Abstraction'' introduced by Longley and 
 Plotkin in \cite{LP96}, as shown \textit{loc. cit}.
 
 The axiomatic methods developed in the present paper can be extended 
 to yield this stronger result. We briefly sketch the necessary 
 extensions. Firstly, we take our sequential categories to be 
 enriched over \emph{enumerated sets} \cite{AL} rather than 
 just pointed sets. All the isomorphisms required in the axioms 
 have then to be given effectively. This leads to an effective version 
 of the Decomposition Theorem as in \cite{AJM}. The development then 
 proceeds exactly as in \cite{AJM}. That is, universal terms are 
 defined in PCF, from which the definability of all strategies follows 
 directly.

\paragraph{Ackowledgements}
The research described in this paper was supported by the U.K. EPSRC grant
``Foundational Structures for Computing Science''.
I am grateful to Guy McCusker and the two anonymous referees for their
comments on the preliminary version of this paper.
 	\bibliography{biblio}
\end{document}